\documentclass[conference]{IEEEtran}

\usepackage{amsfonts}
\usepackage{amssymb}
\usepackage{stfloats}
\usepackage{cite}
\usepackage{graphicx}
\usepackage{psfrag}
\usepackage{subfigure}
\usepackage{amsmath}
\usepackage{array}
\usepackage{algorithm}
\usepackage{algpseudocode}
\usepackage{setspace}
\usepackage[english]{babel}
\usepackage{blindtext}
\usepackage{url}
\usepackage{epstopdf}
\usepackage{verbatim}
\usepackage{color}
\usepackage{amsmath}
\usepackage{todonotes}
\usepackage{booktabs}

\newtheorem{Prob}{Problem}

\title{Channel Estimation for WiFi Prototype Systems with Super-Resolution Image Recovery}

\author{Qi Shi$^{\dagger}$, Yangyu Liu$^{\dagger}$, Shunqing Zhang$^{\dagger}$, Shugong Xu$^{\dagger}$,  Shan Cao$^{\dagger}$, and Vincent LAU$^{\ddag}$\\
$^{\dagger}$ Shanghai Institute for Advanced Communication and Data Science, \\
Key laboratory of Specialty Fiber Optics and Optical Access Networks, \\
Shanghai University, Shanghai, 200444, China\\
$^{\ddag}$ Department of ECE, Hong Kong University of Science and Technology, \\
Clear Water Bay, Kowloon, Hong Kong S. A. R., China \\
Email:\{qi\_shi, yangyuliu, shugong, shunqing, cshan\}@shu.edu.cn, eeknlau@ust.hk}

\begin{document}

\maketitle

\begin{abstract}
Channel estimation is crucial for modern WiFi system and becomes more and more challenging with the growth of user throughput in multiple input multiple output configuration. Plenty of literature spends great efforts in improving the estimation accuracy, while the interpolation schemes are overlooked. To deal with this challenge, we exploit the super-resolution image recovery scheme to model the non-linear interpolation mechanisms without pre-assumed channel characteristics in this paper. To make it more practical, we offline generate numerical channel coefficients according to the statistical channel models to train the neural networks, and directly apply them in some practical WiFi prototype systems. As shown in this paper, the proposed super-resolution based channel estimation scheme can outperform the conventional approaches in both LOS and NLOS scenarios, which we believe can significantly change the current channel estimation method in the near future.
\end{abstract}


\section{Introduction} \label{sect:intro}
Channel estimation is often regarded as a critical component for modern wireless systems \cite{xie2016overview}. By inserting pre-known pilot sequences along with the transmit data, receivers shall be able to estimate the real-time wireless environment accordingly and perform coherent detection thereafter. Since the pilots consume air interface resources without useful information delivery, the existing literature spends tremendous efforts \cite{mezghani2018blind} in improving channel estimation accuracy, especially for multiple input multiple output (MIMO) orthogonal frequency division multiplexing (OFDM) configurations, which is commonly adopted in current cellular or wireless local area networks (WLAN) systems.

Along with the explosive growth of user throughput requirement for future wireless systems, the dimensions of available channels become huge when massive antennas and frequency bands have been aggregated and the accurate channel estimation under high dimensional signal spaces becomes much more challenging \cite{xie2018channel}. Several interesting schemes have been proposed to address this issue. For example, by recognizing the sparsity of MIMO channel characteristics, \cite{rao2014distributed} proposes a distributed compressive sensing scheme with a few pilots for channel state recovery in MIMO systems. While for the case of broadband communication, \cite{dongming2003channel} develops the matching pursuit based channel estimator which is an efficient method for the problem of multipath propagation. In \cite{gao2015spatially}, the authors utilize the non-orthogonal concept to aggregate multiple pilots together and improve the potential resolution of high dimension channel spaces. Nevertheless, the above schemes rely on abstracted channel models with certain characteristics, such as channel sparsity, and the application to practical systems is still challenging， especially when the hardware imperfectness has been considered \cite{cheng2013wideband}.


In addition to the model based channel estimation schemes, the model-free channel estimation approaches, together with recent development of deep learning technologies, have been considered for practical wireless communication systems. For example, \cite{ye2018power} utilizes deep neural networks (DNN) to combine the channel estimation and signal detection procedures in the conventional OFDM systems. In \cite{he2018deep}, a learned denoising-based approximate message passing (LDAMP) neural network has been proposed to reduce the noise effect in the iterative sparse channel estimation processes. The above methods provide insightful results on facilitating the deep learning technology for wireless application, which combines with several wireless domain knowledge, such as channel statistics or noise distributions. In this paper, however, we consider the domain knowledge from the image recovery area, where the deep learning technology has been successfully applied. To be more specific, super-resolution (SR), a deep-learning based image reconstruction technique, is proposed to recover the channel state information (CSI) from the limited observations of pilot signals and the main contributions of this paper are summarized as follows.
\begin{itemize}
    \item{\em Channel State Recovery versus Interpolation} The existing literature for channel estimation focuses on the channel state recovery, which obtains CSI knowledge from the received pilot signals \cite{xie2016overview}, while the interpolation schemes to calculate CSI values at wireless data transmission areas receive limited research attention. This is partially because the non-linear interpolation relation is challenging to characterize and the real-time requirement of channel estimation prevents complicated interpolation algorithms. In this paper, we exploit the SR image recovery scheme to model the non-linear interpolation relations in the conventional channel estimation problems, which does not rely on any pre-assumed statistical channel knowledge.
    \item{\em Numerical versus Practical} In the practical wireless systems, the true CSI knowledge can be hardly obtained. To make the proposed SR based channel estimation approach feasible, we offline generate some numerical channel coefficients according to the statistical channel models to train the SR neural networks, and then apply them online to some practical WiFi prototype systems. As we will show later, the proposed SR based channel estimation scheme can outperform several baseline schemes, which can provide some design insights for future deep learning based wireless communication system.
\end{itemize}

The rest of the paper is organized as follows. In Section~\ref{sect:sys_model}, we introduce the widely used MIMO-OFDM system model and summarize the traditional channel estimation as well as interpolation schemes. A brief formulation using SR framework for channel estimation is proposed in Section~\ref{sect:map} and the associated neural network is analyzed in Section~\ref{sect:SR_comp}. In Section~\ref{sect:simu}, we verify the estimation accuracy via practical measurements by comparing the proposed SR based channel estimation scheme with conventional baselines and final remarks are given in Section~\ref{sect:conc}.

{\em Notations}: Boldface uppercase and lowercase denote matrices and vectors, respectively, and the matrix inverse and conjugate transpose operations are given by $(\cdot)^{-1}$ and $(\cdot)^H$.  $\mathbb{E}\left[\cdot\right]$ defines the mathematical expectation, while the matrix Frobenius is given by $\|\mathrm{\cdot}\|_F$.

\section{System Model and Conventional Schemes} \label{sect:sys_model}
In this section, we introduce the mathematical model for general MIMO-OFDM systems and explain the conventional channel state recovery and interpolation mechanisms in details.

\subsection{System Model}\label{subsect:syslmodel}
\begin{figure}[htbp]
\centering
\subfigure[Real Deployment]{
\begin{minipage}[t]{0.5\linewidth}
\centering
\includegraphics[width=1.5in]{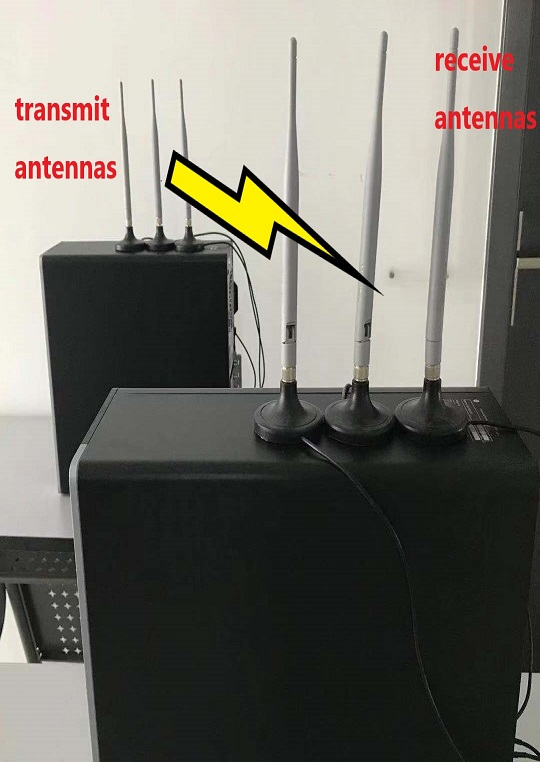}
\end{minipage}%
}%
\subfigure[System Model]{
\begin{minipage}[t]{0.5\linewidth}
\centering
\includegraphics[width=1.5in]{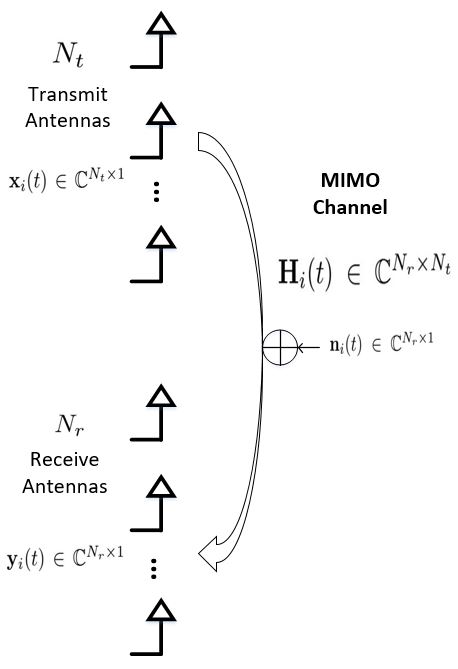}
\end{minipage}%
}%
\centering
\caption{An overview of WiFi prototype systems with $N_{r} \times N_{t}$ antenna configurations. Sub-figure (a) shows the real deployment of WiFi prototype systems and the abstracted model is shown in sub-figure (b). The MIMO channel conditions between transmit and receive antenna pairs for subcarrier $i$ are denoted as $\mathbf{H}_{i}(t)$.}
\label{fig:sysmodel}
\end{figure}

Consider a typical MIMO-OFDM system, as is depicted in Fig.~\ref{fig:sysmodel}, with $N_t$ transmit and $N_r$ receive antennas in the wireless fading environment. The received symbols at the $i^{th}$ subcarrier, $\mathbf{y}_{i}(t) \in \mathbb{C}^{N_r \times 1}$, after the Fast Fourier Transform (FFT) processing, can be modeled through,
\begin{eqnarray} \label{eqn:awgn}
\mathbf{y}_{i}(t) = \mathbf{H}_{i}(t)\mathbf{x}_{i}(t) + \mathbf{n}_{i}(t),
\end{eqnarray}
where $\mathbf{H}_{i}(t) \in \mathbb{C}^{N_r \times N_t}$ denotes the MIMO fading coefficients, $\mathbf{x}_{i}(t)  \in \mathbb{C}^{N_t \times 1}$ denotes the transmitted symbols, and $\mathbf{n}_{i}(t) \in \mathbb{C}^{N_r \times 1}$ denotes the additive white Gaussian noise (AWGN) with zero mean and unit variance. Based on the observed $\mathbf{y}_{i}(t)$ and the pre-known pilot sequences $\mathbf{x}_{i}(t)$, we can recover the channel coefficients $\mathbf{H}_{i}(t)$ accordingly.

\begin{figure}[t]
\centering
\subfigure{
  \includegraphics[width=2.8in]{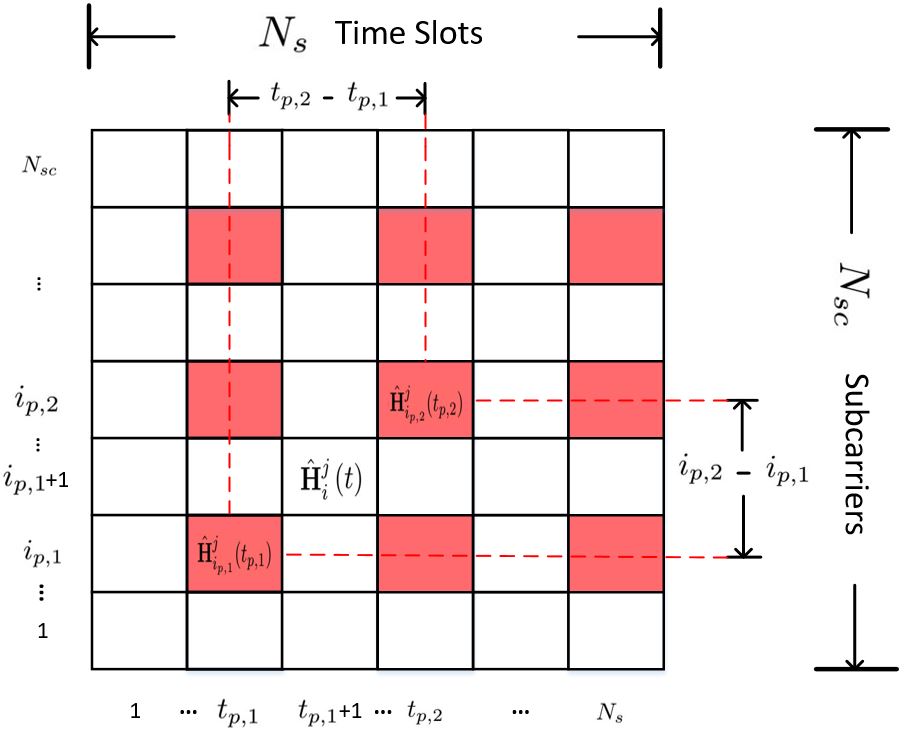}}
\hfill
\caption{Illustration of interpolation mechanisms within one RB, containing $N_{sc}$ subcarriers in the frequency domain and $N_s$ slots in the time domain. Red grids represent the pilot locations, and the channel conditions at white grids are estimated through interpolation mechanisms defined in $\eqref{eqn:LI}$ and $\eqref{eqn:GI}$.}
\label{fig:channelmatrix}
\end{figure}

In the practical system, to minimize the channel estimation overhead, the entire process is performed on a resource block (RB) basis (usually in accordance with the coherence time and coherence bandwidth) with $N_s$ time slots and $N_{sc}$ subcarriers as shown in Fig.~\ref{fig:channelmatrix}. Denote $\mathbf{H}_{i}^{j}(t)$ to be the MIMO fading coefficients for the $j^{th}$ RB, and by arranging them in order, we can obtain the entire channel condition, $\mathbf{H}^{j}$, as follows,
\begin{eqnarray}
\mathbf{H}^{j} & = & \{\mathbf{H}_{i}^{j}(t)\}, \nonumber \\ & = &
\left[
\begin{matrix}
 \mathbf{H}_{1}^{j}(1)      & \mathbf{H}_{1}^{j}(2)      & \cdots & \mathbf{H}_{1}^{j}(N_{s})      \\
 \mathbf{H}_{2}^{j}(1)      & \mathbf{H}_{2}^{j}(2)      & \cdots & \mathbf{H}_{2}^{j}(N_{s})      \\
 \vdots & \vdots & \ddots & \vdots \\
 \mathbf{H}_{N_{sc}}^{j}(1)      & \mathbf{H}_{N_{sc}}^{j}(2)      & \cdots & \mathbf{H}_{N_{sc}}^{j}(N_{s})      \\
\end{matrix}
\right].
\end{eqnarray}
In addition, to control the resource usage for channel estimation, only limited locations are selected for pilot allocation. Denote $\Omega_{p}$ to be the collections of pilot positions and the aggregated CSIs for pilots can be expressed through,
\begin{eqnarray}
\mathbf{H}^{j}_{p} = \left\{\mathbf{H}_{i_p}^{j}(t_p), \forall (i_p,t_p) \in \Omega_{p} \right\},
\end{eqnarray}
where $i_p$ and $t_p$ denote the indexes of subcarrier and time slot within RB.

\subsection{Conventional Schemes}\label{subsect:conventionalmodel}
In the conventional channel estimation process, receivers perform channel state recovery to obtain an estimated channel conditions at the pilot positions, $\hat{\mathbf{H}}^{j}_{p} = \{\hat{\mathbf{H}}_{i_p}^{j}(t_p), \forall (i_p,t_p) \in \Omega_{p}\}$, and then apply interpolation mechanisms to get the entire channel coefficients, $\hat{\mathbf{H}}^{j} = \{\hat{\mathbf{H}}_{i}^{j}(t)\}$.

\subsubsection{Channel State Recovery}\label{subsubsect:convenCSI}
As mentioned before, the channel state recovery aims to obtain CSI knowledge at the pilot locations. For all the possible $(i_p,t_p) \in \Omega_{p}$, we can rewrite \eqref{eqn:awgn} as follows,
\begin{eqnarray} \label{eqn:awgn_pilot}
\mathbf{y}^{j}_{i_p}(t_p) = \mathbf{H}^{j}_{i_p}(t_p)\mathbf{x}_{i_p}(t_p) + \mathbf{n}^{j}_{i_p}(t_p),
\end{eqnarray}
where we omit the superscript $j$ for $\mathbf{x}_{i_p}(t_p)$ as the same pilot sequences are adopted for different RBs in practical systems. With the above relation, the classical least square (LS) and minimum mean square error (MMSE) \cite{van1995channel} estimation can be described through,
\begin{eqnarray}
\hat{\mathbf{H}}_{i_p}^{j,\textrm{LS}}(t_p) & = & \mathbf{y}^{j}_{i_p}(t_p)\mathbf{x}_{i_p}^H(t_p)\left(\mathbf{x}_{i_p}(t_p)\mathbf{x}_{i_p}^H(t_p)\right)^{-1}, \label{eqn:ls}\\
\hat{\mathbf{H}}_{i_p}^{j,\textrm{MMSE}}(t_p) & = & \mathbf{R}_{\mathbf{H}}\left[\mathbf{R}_{\mathbf{H}} + \left(\mathbf{x}_{i_p}(t_p)\mathbf{x}_{i_p}^H(t_p)
\right)^{-1}\right]^{-1} \nonumber \\
&& \hat{\mathbf{H}}_{i_p}^{j,\textrm{LS}}(t_p), \label{eqn:mmse}
\end{eqnarray}
where the unit variance assumption for the additive noise $\mathbf{n}^{j}_{i_p}(t_p)$ is applied and $\mathbf{R}_{\mathbf{H}}$ is the channel correlation matrix at the receiver side with $\mathbf{R}_{\mathbf{H}}=\mathbb{E}\left[\mathbf{H}_{i_p}(t)\mathbf{H}_{i_p}^H(t)\right]$.

\subsubsection{Interpolation Mechanisms}\label{subsubsect:interpolation}
With the estimated channel states at pilot positions, $\hat{\mathbf{H}}^{j}_{p}$, the interpolation mechanisms target to find the function $G(\cdot)$ that can minimize average mean square estimation error of $\hat{\mathbf{H}}^{j}$. Mathematically, the interpolation process can be modeled as,
\begin{eqnarray}\label{eqn:min}
\begin{aligned} G^{\star}(\cdot) = & \mathop{\arg\min}_{G(\cdot)} & \lim_{J\rightarrow \infty} \frac{1}{J} \sum_{j=1}^{J}\|\mathbf{H}^{j}-\hat{\mathbf{H}}^{j}\|_{F}^2, \\
& \textrm{subject to} 
& \hat{\mathbf{H}}^{j} = G\left(\hat{\mathbf{H}}^{j}_{p}\right). \qquad
\end{aligned}
\end{eqnarray}
In the practical deployment, since the real channel responses $\mathbf{H}^{j}$ is difficult to obtain, the above minimization is in general difficult to solve. Existing approaches rely on some heuristic interpolation schemes, such as linear interpolation (LI) and Gaussian interpolation (GI) \cite{huaqing2010two}. Specifically, given two neighboring pilot locations $(i_{p,1},t_{p,1})$ and $(i_{p,2},t_{p,2})$, for any $t_{p,1} \leq t \leq t_{p,2}$ and $i_{p,1} \leq i \leq i_{p,2}$, we have the linear and Gaussian interpolated results $\hat{\mathbf{H}}_{i,LI}^{j}(t)$ and $\hat{\mathbf{H}}_{i,GI}^{j}(t)$ as follows,

\begin{align}
&\hat{\mathbf{H}}_{i,LI}^{j}(t)  = G_{LI}\left(\hat{\mathbf{H}}^{j}_{p}\right)\nonumber= \left(1-\alpha\right)\left(1-\beta\right)\\
&\times\hat{\mathbf{H}}_{i_{p,1}}^{j}(t_{p,1})+\alpha\beta\hat{\mathbf{H}}_{i_{p,2}}^{j}(t_{p,2}),\label{eqn:LI} \\
&\hat{\mathbf{H}}_{i,GI}^{j}(t) = G_{GI}\left(\hat{\mathbf{H}}^{j}_{p}\right)= \frac{1}{4}\left(\alpha^2-\alpha\right)\left(\beta^2-\beta\right) \nonumber\\
&\times\hat{\mathbf{H}}_{2i_{p,1}-i_{p,2}}^{j}(2t_{p,1}-t_{p,2})+\left(1-\alpha^2\right)\left(1-\beta^2\right) \hat{\mathbf{H}}_{i_p}^{j}(t_p)\nonumber\\
&+\frac{1}{4}\left(\alpha^2+\alpha\right)\left(\beta^2+\beta\right)\hat{\mathbf{H}}_{2i_{p,1}+i_{p,2}}^{j}(2t_{p,1}+t_{p,2}),\label{eqn:GI}
\end{align}
where $\alpha=\frac{i-i_{p,1}}{i_{p,2}-i_{p,1}}$ and $\beta=\frac{t-t_{p,1}}{t_{p,2}-t_{p,1}}$ denote the interpolation coefficients.

Although the channel state recovery and interpolation provides a useful solution to the channel estimation for MIMO-OFDM systems, the achievable estimation accuracy is limited due to the following reasons. First of all, the time/frequency domain channel responses are never flat within each RB and the LI or GI can not well approximate the nonlinear variations in the practical scenarios. Second, the spatial correlations among different antennas are ignored in the current approach, which can be exploited via more complicated functions of $G(\cdot)$. Last but not least, the estimation performance can be further enhanced via a joint optimization for channel state recovery and interpolation. 

\section{Super-Resolution Framework} \label{sect:map}
In this section, we propose a novel interpolation scheme by applying the deep learning based SR image reconstruction technique, and elaborate the generation of data sets as well as the candidate neural networks. 
\subsection{SR Formulation} \label{subsect:imgmodel}
\begin{figure}[t]
\centering
\includegraphics[width = 2.8 in]{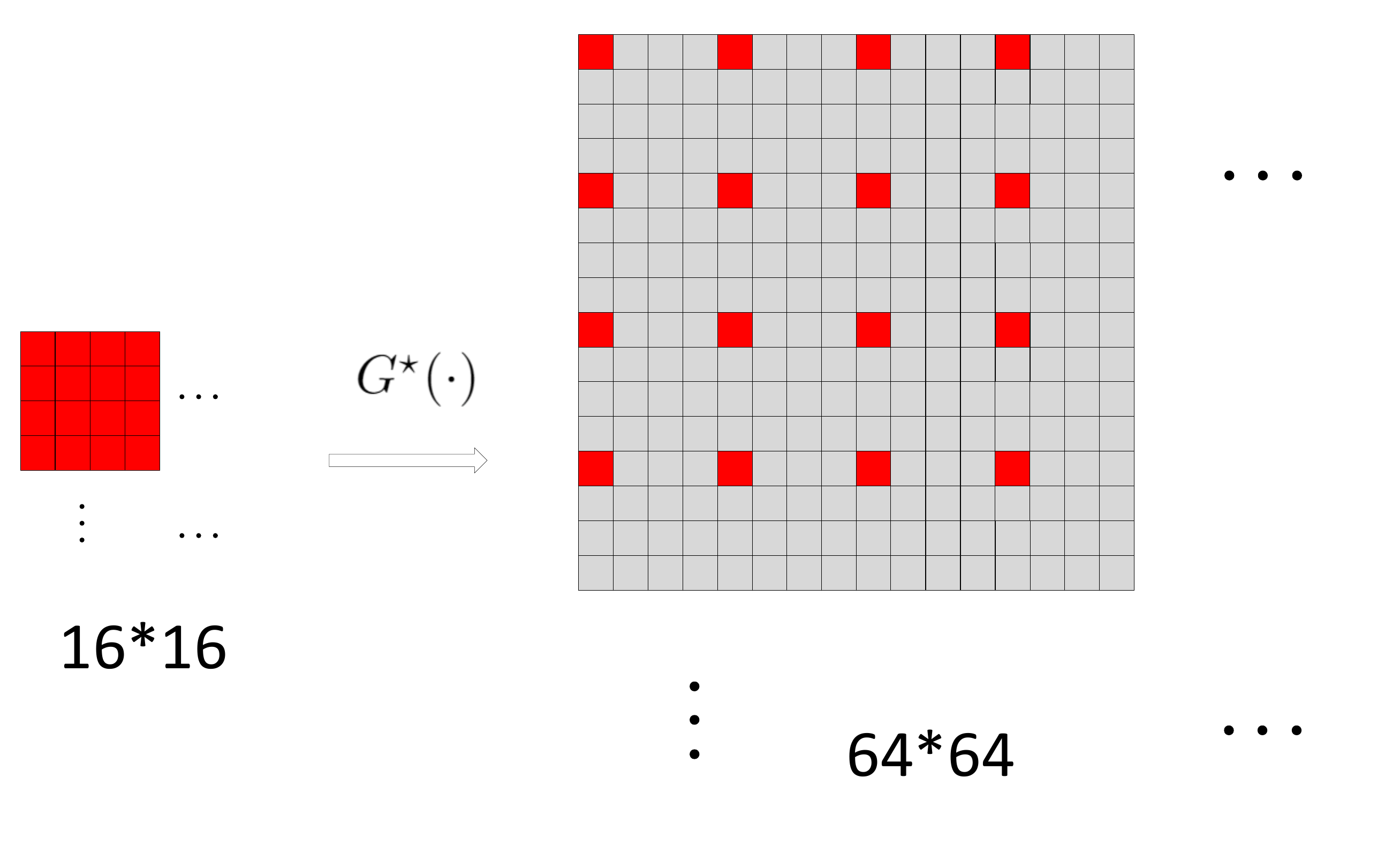}
\hfill
\caption{Illustration of optimal interpolation function $G^{\star}(\cdot)$, where the channel conditions estimated at the pilot locations ($16 \times 16$ red grids) needs to be converted into $64 \times 64$ grids via better interpolation mechanisms.}
\label{fig:pilot}
\end{figure}

Since the number of pilots, $N_{p}$, is always less than the number of resources, $N_{sc} \times N_{s}$,  in each RB, the estimated channel conditions $\hat{\mathbf{H}}_{p}^{j}$ and $\hat{\mathbf{H}}^{j}$ can be regarded as low-resolution (LR) and high-resolution (HR) images with $N_{p}$ and $N_{sc} \times N_{s}$ pixels, respectively. Therefore, the function $G(\cdot)$ is a typical SR operation in the image processing area with the RGB color information determined by $N_{R} \times N_{T}$ complex numbers. With the above understanding, the SR framework as defined in \cite{Dong2014Learning} can be directly applied to approximate the optimal $G^{\star}(\cdot)$ as defined in \eqref{eqn:min}, whose primary function is illustrated in Fig.~\ref{fig:pilot}.

As the objective function in \eqref{eqn:min} is the mean square errors (MSE) between the true channel condition $\mathbf{H}^{j}$ and the estimated channel condition $\hat{\mathbf{H}}^{j}$, we can construct the loss function in the deep learning based SR framework as,
\begin{eqnarray}
\mathcal{L}\left(\hat{\mathbf{H}}^{j}\right) = \frac{1}{J} \sum_{j=1}^{J}\|\mathbf{H}^{j}-\hat{\mathbf{H}}^{j}\|_{F}^2,
\end{eqnarray}
where the actual channel conditions $\mathbf{H}^{j}$ can be regarded as the ground truth HR image.

\subsection{Data Sets}
Different from the traditional training data sets generation in the previous SR research, where a HR image is given in advance and the associated LR image is obtained through down-sampling, the true channel conditions $\mathbf{H}^{j}$ are difficult to collect in general. To overcome this obstacle, we generate $\mathbf{H}^{j}$ through a statistical channel model, e.g. COST 2100 as defined in \cite{liu2012cost}, and simulate the pilot transmission process as defined in \eqref{eqn:awgn_pilot} via numerical examples. Based on the classical LS and MMSE estimation methods, e.g. \eqref{eqn:ls} and \eqref{eqn:mmse}, we generate LR images $\hat{\mathbf{H}}^{j,\textrm{LS}}_{p} = \left\{\hat{\mathbf{H}}^{j,\textrm{LS}}_{i_{p}}(t_{p})\right\}$ and $\hat{\mathbf{H}}^{j,\textrm{MMSE}}_{p} = \left\{\hat{\mathbf{H}}^{j,\textrm{MMSE}}_{i_{p}}(t_{p})\right\}$, respectively. 

Through this approach, we can generate sufficient large size of data sets for training and evaluation, and to accommodate with the practical WiFi prototype system configuration, we choose $N_{r} = N_{t} = 3$, $N_{sc} = 64$, and $N_{s} = 64$ in the SR neural network evaluation and optimization.

\subsection{Candidate Neural Networks}
With the generated data sets and the clearly defined objective, we shall figure out the suitable neural network architecture for this type of application. Based on the existing literature, two types of neural networks show superior restoration performance over the traditional SR technology, which are summarized as follows.

\begin{itemize}
    \item {\em SR-CNN}\cite{Dong2014Learning}: The LR image, $\hat{\mathbf{H}}_{p}^{j}$, is expanded to the HR image, $\hat{\mathbf{H}}_{LI}^{j}= \{\hat{\mathbf{H}}_{i,LI}^{j}(t)\}$ or $\hat{\mathbf{H}}_{GI}^{j}= \{\hat{\mathbf{H}}_{i,GI}^{j}(t)\}$, in the first stage, and then using neural networks to approximate the non-linear relation between $\hat{\mathbf{H}}_{LI}^{j} / \hat{\mathbf{H}}_{GI}^{j}$ and $\hat{\mathbf{H}}^{j}$.
    \item{\em EDSR}\cite{Lim2017Enhanced}: Apply a residual learning network \cite{DBLP:journals/corr/HeZRS15} to progressively predict the SR images, $\hat{\mathbf{H}}^{j}$, from LR images, $\hat{\mathbf{H}}_{p}^{j}$.
\end{itemize}

Although the aforementioned deep learning based SR schemes achieves satisfied peak signal to noise ratio (PSNR) performance on the public data sets, such as BSD100 \cite{arbelaez2011contour}, the corresponding behaviors for approximating the optimal interpolation function $G^{\star}(\cdot)$ is still unknown. 
\section{SR Framework Analysis} \label{sect:SR_comp}
In this section, we focus on analyzing the SR neural networks by comparing PSNR performance under the statistical COST 2100 channel models and exploits the possibility of applying SR based method for CSI estimation.

\subsection{System Architecture}\label{subsect:sysArch}
\begin{figure}
\centering
\subfigure[SR-CNN]{
\includegraphics[width=2 in]{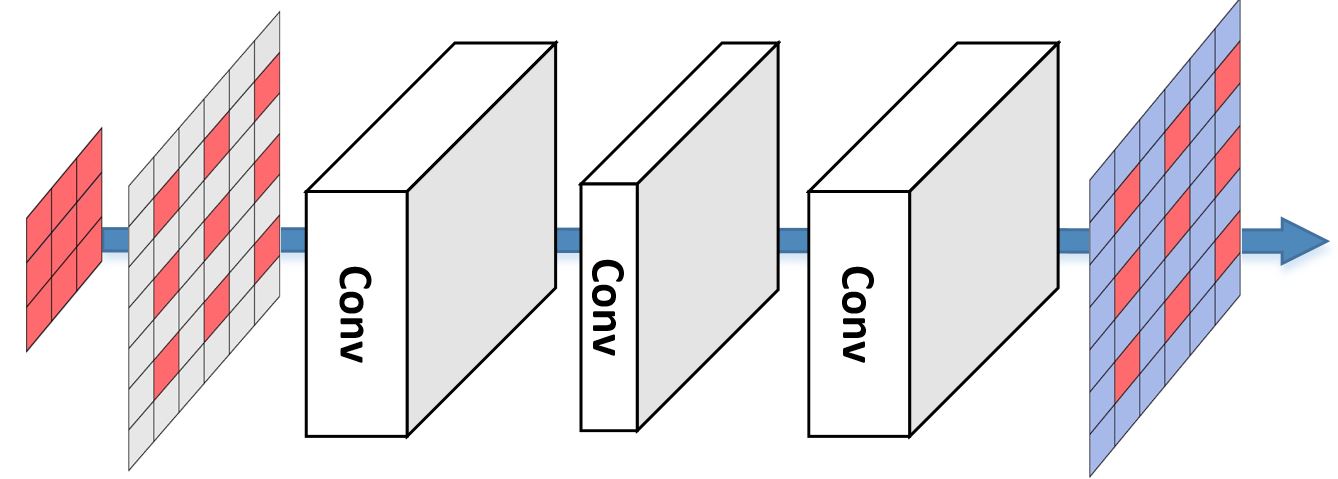}}
\hfill
\centering
\subfigure[EDSR]{
\includegraphics[width=2.5 in]{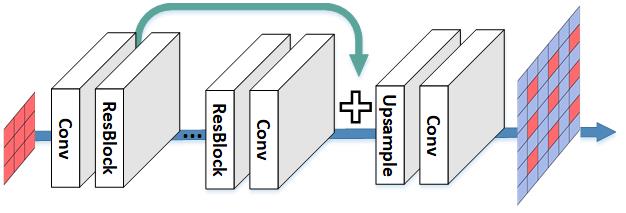}}
\caption{The network architecture of two typical SR neural networks, where sub-figure (a) and (b) demonstrates the abstracted network configuration of SR-CNN and EDSR, respectively. Different from the traditional linear interpolation or Gaussian interpolation, SR-CNN and EDSR approximate the non-linear interpolation via neural networks.}
\label{fig:model_SR}
\end{figure}

The architecture of SR-CNN and EDSR can be found in \cite{Dong2014Learning} and \cite{Lim2017Enhanced}. As is shown in Fig.~\ref{fig:model_SR}, 
SR-CNN models the non-linear relationship between the LI$/$GI result and the real-time captured result, while EDSR applies massive residual network blocks to gradually generate the fine-grained result. Mathematically, the SR-CNN and EDSR based formulations are given as follows.
\begin{Prob}[SR-CNN Formulation]
In the SR-CNN formulation, the function $G^{\star}(\cdot)$ is approximated through $G^{\star}(\cdot) \approx G^{\star}_{CNN}(G_{LI/GI}(\cdot))$, with $G^{\star}_{CNN}(\cdot)$ given by
\begin{eqnarray}
G^{\star}_{CNN}(\cdot) & = & \mathop{\arg\min}_{G_{CNN}(\cdot)}\lim_{J\rightarrow \infty} \frac{1}{J} \sum_{j=1}^{J}\|\mathbf{H}^{j}-\hat{\mathbf{H}}^{j}\|_{F},\nonumber\\
& & \textrm{subject to} \ \hat{\mathbf{H}}^{j} = G_{CNN}\left(G_{LI/GI}\left(\hat{\mathbf{H}}^{j}_{p}\right)\right).\nonumber
\end{eqnarray}
\end{Prob}

\begin{Prob}[EDSR Formulation] In the EDSR formulation, we directly approximate $G^{\star}(\cdot)$ via progressively minimize the difference between the estimated results $\hat{\mathbf{H}}^{j}$ and the real channel state $\mathbf{H}^{j}$.
\begin{eqnarray}
G^{\star}(\cdot) &\approx& G^{\star}_{EDSR}(\cdot)\nonumber \\
& = & \mathop{\arg\min}_{G_{EDSR}(\cdot)}  \lim_{J\rightarrow \infty} \frac{1}{J} \sum_{j=1}^{J}\|\mathbf{H}^{j}-\hat{\mathbf{H}}^{j}\|_{1},\nonumber \\
&& \textrm{subject to}\ \ \hat{\mathbf{H}}^{j} = G_{EDSR}\left(\hat{\mathbf{H}}^{j}_{p}\right).
\end{eqnarray}
where $\|\cdot\|_{1}$ denotes the element-wise absolute value summation of the inner matrix.
\end{Prob}

From the above formulations, since SR-CNN
focuses on the non-linear relationship between linear or Gaussian interpolated channel conditions and the real channel measurements, the optimization space for SR-CNN is in general limited. For EDSR, although the objective function is built on $l-1$ norm, the difference between $\|\mathbf{H}^{j}-\hat{\mathbf{H}}^{j}\|_{1}$ and the top of $\|\mathbf{H}^{j}-\hat{\mathbf{H}}^{j}\|_{F}$ is margined if $\|\mathbf{H}^{j}-\hat{\mathbf{H}}^{j}\|_{1}$ is sufficiently small. As a result, EDSR can provide better performance than SR-CNN approach under the same condition, which is verified in the coming subsection.

\subsection{Network Training \& Comparison}\label{subsect:modeltraining}
To generate sufficient data sets for neural network training, we consider the famous COST 2100 model to generate ${\mathbf{H}}^{j}$ as well as ${\mathbf{H}}_p^{j}$ as defined in Section~\ref{sect:sys_model}. To be more specific, the pilot positions are given by $\left\{\Omega_{p}: i_p\in \left\{1,5,9,\cdots,61\right\},\  t_p\in\left\{1,5,9,\cdots,61\right\}\right\}$ as shown in Fig.~\ref{fig:pilot}, and we generate $J = 2000$ frames for the training and 200 frames for the validation test. The validation results are listed in Table \ref{tab:compare_model}, comparing SR networks with conventional channel estimation methods mentioned in Section~\ref{sect:sys_model}, where the PSNR is selected as the performance metric\footnote{PSNR is commonly adopted in the machine learning area for SR recovery, as shown in \cite{Dong2014Learning}}. Based on Table \ref{tab:compare_model}, EDSR takes the advantage of progressive learning for the entire function $G^{\star}(\cdot)$, which co-verifies the analysis in the previous subsection. 

\begin{table}[ht]
\caption{PSNR comparison of SR-CNN \& EDSR based and conventional channel estimation schemes LS \& MMSE}
\label{tab:compare_model}
\centering
\footnotesize
\begin{tabular}{c c}
\toprule
 \textbf{Interpolation Mechanisms}&\textbf{PSNR (LS/MMSE)}\\
\midrule
SR-CNN &25.7762$/$25.8396\\
\midrule
EDSR &30.7883$/$31.0798\\ 
\midrule
Linear Interpolation & 21.0597$/$21.0671 \\ 
\midrule
Gaussian Interpolation & 21.1140$/$21.1150 \\ 
\bottomrule
\end{tabular}
\end{table}


\section{Empirical Results} \label{sect:simu}
In this section, we deploy the SR based channel estimation scheme in the WiFi prototype systems using the trained SR neural network as illustrated in Section~\ref{sect:SR_comp} and provide some empirical results to show the effectiveness of the proposed channel estimation scheme by comparing with other traditional methods summerized in Section~\ref{sect:sys_model}. As shown in Fig.~\ref{fig:sysmodel}(a), two commercial desktops\footnote{In order to emulate the MIMO-OFDM scenario, both two desktops are equipped with Qualcomm Atheros AR9380 WiFi netwowrk interface cards, and on top of that, a open source tool Atheros CSI Tool \cite{Xie2015Precise} is installed to obtain the real time channel coefficients.} are communicating with each other through IEEE 802.11n  WiFi protocols \cite{5307322}.  
Since some of the subcarriers are muted as guard subcarriers in WiFi prototype systems, we slightly modify the input and output matrices, where $N_{sc} = N_{s} = 56$ is selected and the number of pilots are scaled accordingly. The detailed parameters adopted in the evaluation are summarized in Table~\ref{tab:para_evalu}.
\begin{table}[h]
\caption{Related Parameters Used in Evaluation}
\label{tab:para_evalu}
\centering
\footnotesize
\begin{tabular}{c c c c c}
\toprule
\textbf{Scenario} & COST 2100&WiFi Prototype\\
\midrule
\textbf{Frequency} &5.3 GHz&5.32 GHz  \\ 
\midrule
\textbf{Bandwidth} &20 MHz & 20 MHz  \\ 
\midrule
\textbf{Antenna Configuration} &3 $\times$ 3 &  3 $\times$ 3\\ 
\midrule
\textbf{Number of Packets} &128000 & 60000 \\ 
\bottomrule
\end{tabular}
\end{table}


\subsection{LOS Scenario}
\begin{figure}[t]
\centering
\includegraphics[width = 3.4 in]{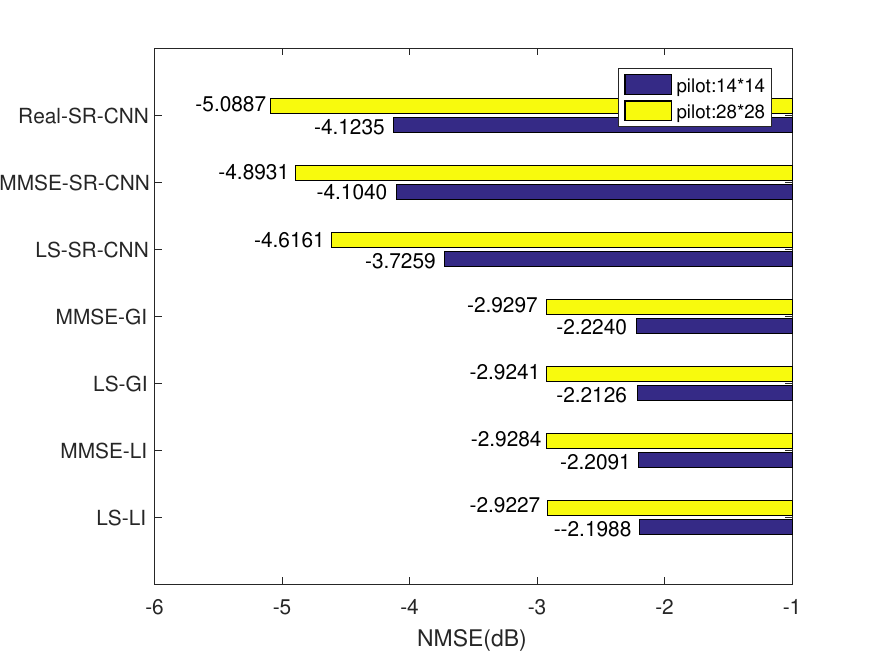}
\hfill
\caption{NMSE comparison of SR-CNN based and traditional channel estimation schemes in LOS environments. Two types of pilot arrangements are considered and the received SNR is fixed to 20 dB. As shown in this figure, the SR-CNN based channel estimation scheme outperforms the traditional LI and GI approaches for both two pilot arrangement cases. }
\label{fig:LOS}
\end{figure}

In the first experiment, we compare the SR-CNN based channel estimation method with conventional schemes to show the effectiveness of the proposed SR based framework. LS and MMSE algorithms are adopted for channel state recovery at the pilot locations and we test two different pilot arrangements with $14 \times 14$ and $28 \times 28$ configurations. As is shown in Fig.~\ref{fig:LOS}, the SR-CNN based interpolation scheme outperforms the traditional LI or GI approach under both LS and MMSE channel state recovery methods, where the normalized MSE (NMSE)  performance reduces from more than $-3$ dB to $-4$ dB. In addition, by comparing Real-SR-CNN and MMSE-SR-CNN curves, we can show that the SR networks trained from COST 2100 model is already sufficient for channel estimation in the practical system, and the NMSE loss is less than $0.2$ dB for $28 \times 28$ case and $0.02$ dB for $14 \times 14$ cases.

\subsection{NLOS Scenario}
\begin{figure}[t]
\centering
\includegraphics[width = 3.4 in]{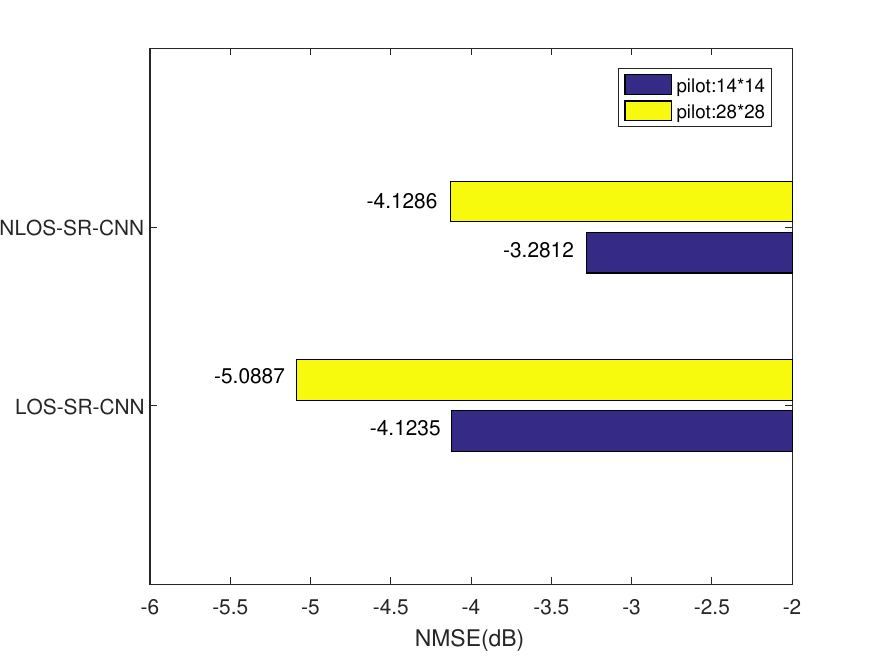}
\hfill
\caption{NMSE comparison of SR-CNN based and traditional channel estimation schemes in NLOS environments. Two types of pilot arrangements are considered and the received SNR is fixed to 20 dB. As shown in this figure, the SR based interpolation scheme achieves a reliable NMSE performance in NLOS conditions.}
\label{fig:NLOS}
\end{figure}

In the second experiment, we extend the above experiments into NLOS scenario to show the robustness of the proposed algorithms in Fig.~\ref{fig:NLOS}. Due to the potential scattering effects, the achievable NMSE will be degraded by $0.9$ dB for NLOS scenario. As the channel estimation in the NLOS environment is in general more difficult than LOS case as elaborated in \cite{benedetto2007dynamic}, the proposed SR based channel estimation scheme achieves a reliable NMSE performance, which can be extended to NLOS application as well.

\subsection{Performance under EDSR}
\begin{figure}[t]
\centering
\includegraphics[width = 3.4 in]{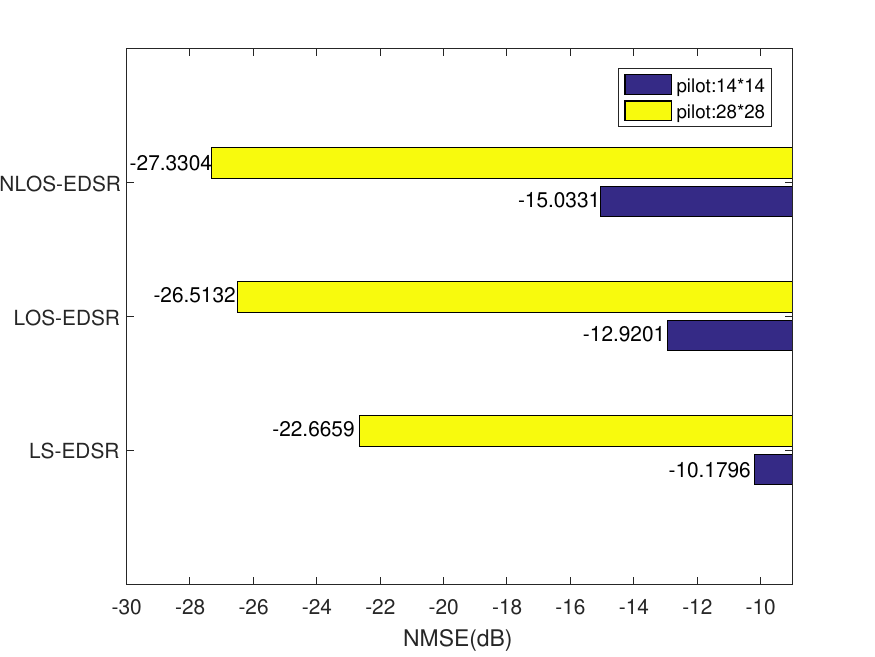}
\hfill
\caption{NMSE performance of EDSR networks under two different types of pilot arrangements in LOS and NLOS environments. The received SNR is fixed to 20 dB. As shown in this figure, the EDSR based channel estimation scheme achieves better NMSE performance in both LOS and NLOS conditions for two pilot arrangement cases.
}
\label{fig:EDSR_V1}
\end{figure}

In the third experiment, we redo the above experiments by replacing the SR neural networks, e.g. from SR-CNN to EDSR. As EDSR provides better PSNR performance as shown in previous section, the better NMSE performance can be expected. As shown in Fig.~\ref{fig:EDSR_V1}, for both LOS and NLOS scenarios, EDSR based schemes achieve less than $-10$ dB NMSE for $14 \times 14$ configuration and less than $-20$ dB NMSE for $28 \times 28$ configuration. This is partially because the channel conditions are more close to sparse images and the gradually learning approach can adopt to the tiny variations.


\section{Conclusion} \label{sect:conc}
In this paper, we propose a novel channel estimation method for WiFi prototype systems using SR based image recovery method. To make the proposed scheme more practical, we train the neural networks via numerical simulation results generated from COST 2100 model and directly apply it to the practical systems. Through the empirical examples, we show that the proposed SR based channel estimation methods provide significant performance gains in terms of NMSE over the traditional schemes in the prototype systems, which eventually benefits the channel estimation design for future wireless systems.  

\bibliographystyle{IEEEtran}
\bibliography{IEEEabrv,bb_rf}

\end{document}